\documentclass[prl,aps,superscriptaddress, twocolumn]{revtex4-2}

\usepackage{graphicx,bm}
\usepackage{amsmath}
\usepackage{amssymb}
\usepackage{braket}
\usepackage{epsfig}
\usepackage{epsf}
\usepackage[usenames]{color}

\newcommand{\kB}{{\ensuremath{k_{\mathrm{B}}}}}

\begin{document}

\title{Entropic magnetic interlayer coupling}

\author{W.J. Huddie}

\affiliation{Institute for Theoretical Physics, Utrecht
	University, Leuvenlaan 4, 3584 CE Utrecht, The Netherlands}
\affiliation{Department of Applied Physics, Eindhoven University of Technology, P.O. Box 513, 5600 MB Eindhoven, The Netherlands}

\author{Laura Filion}

\affiliation{Soft Condensed Matter and Biophysics, Debye Institute for Nanomaterials Science, Utrecht University, Utrecht, Netherlands}

\author{Marjolein Dijkstra}

\affiliation{Soft Condensed Matter and Biophysics, Debye Institute for Nanomaterials Science, Utrecht University, Utrecht, Netherlands}

\author{R. A. Duine}

\affiliation{Institute for Theoretical Physics, Utrecht
University, Leuvenlaan 4, 3584 CE Utrecht, The Netherlands}

\affiliation{Department of Applied Physics, Eindhoven University of Technology, P.O. Box 513, 5600 MB Eindhoven, The Netherlands}

\date{\today}

\begin{abstract}
Nanomagnetism concerns the engineering of magnetic interactions in heterostructures that consist of layers of magnetic and non-magnetic materials. Mostly, these interactions are dominated by the minimization of energy. Here, we propose an effective magnetic interlayer coupling that is dominated by the maximization of entropy. As an example, we consider the system that mediates the effective interactions to be square spin ice, in which case we find purely entropic interactions that are long-ranged. We argue that in the thermodynamic limit the entropic interlayer coupling gives rise to entropic torques on the magnetization direction. For small systems, the physical properties are well characterized by the mutual information between the two magnets that are coupled.  Because entropic interactions become stronger for higher temperatures, our findings may benefit the development of nanomagnetic devices that require thermal stability. 
\end{abstract}


\maketitle

\def\bx{{\bm x}}
\def\bk{{\bm k}}
\def\bK{{\bm K}}
\def\bq{{\bm q}}
\def\br{{\bm r}}
\def\bp{{\bm p}}
\def\bM{{\bm m}}
\def\bs{{\bm s}}
\def\bB{{\bm B}}
\def\bj{{\bm j}}
\def\bF{{\bm F}}
\def\id{{\rm d}}
\def\bE{{\bm E}}
\def\bz{{\bm z}}

\def\br{{\bm r}}
\def\bv{{\bm v}}

\def\half{\frac{1}{2}}
\def\args{(\bm, t)}

{\it Introduction.} --- The goal of nanomagnetism is to engineer functional heterostructures of magnetic and non-magnetic materials for applications \cite{SHINJO20091}. The wide range of materials --- and interfaces between them --- allows for a high tunability of the effective interactions between the magnetic layers. A prime example is the Ruderman-Kittel-Kasuya-Yosida (RKKY) interlayer exchange interaction that couples two magnetic layers across a non-magnetic spacer \cite{Grunberg1986,Majkrzak1986,Salamon1986}. This interaction oscillates as a function of the thickness of the non-magnetic spacer, allowing it to be tuned between ferromagnetic and antiferromagnetic couplings, which favor parallel and antiparallel alignment of the magnetic layers, respectively \cite{Slonczewski1989,Parkin1990,Edwards1991,Bruno1995,Faure-Vincent2002}. The antiferromagnetic coupling enables the engineering of magnetic superlattices that demonstrate giant magnetoresistance, a discovery that to a large extent kickstarted nanomagnetism and spintronics \cite{Baibich1988,Binasch1989}. 

Other examples of nanomagnetic interactions are perpendicular magnetic anisotropy \cite{TUDU2017329} and interfacial Dzyaloshinskii-Moriya interactions \cite{PhysRevLett.44.1538}. Both of these arise at an interface between a magnetic material and a heavy metal. The latter provides the strong spin-orbit coupling that, together with the inversion asymmetry provided by the interface, underpins these interactions. Controlling the interfacial Dzyaloshinskii-Moriya interactions is necessary from the point of view of skyrmionics, which aims to use magnetic skyrmions as data carriers in magnetic memories \cite{Back_2020}. The most recent addition to the family of nanomagnetic interactions is the interlayer Dzyaloshinskii-Moriya interaction. This interaction occurs in situations where the non-magnetic spacer connecting two magnetic layers itself breaks inversion symmetry and has sufficiently strong spin-orbit coupling \cite{Fernandez-Pacheco2019,Han2019}.

\begin{figure} 
	\includegraphics[width=9cm]{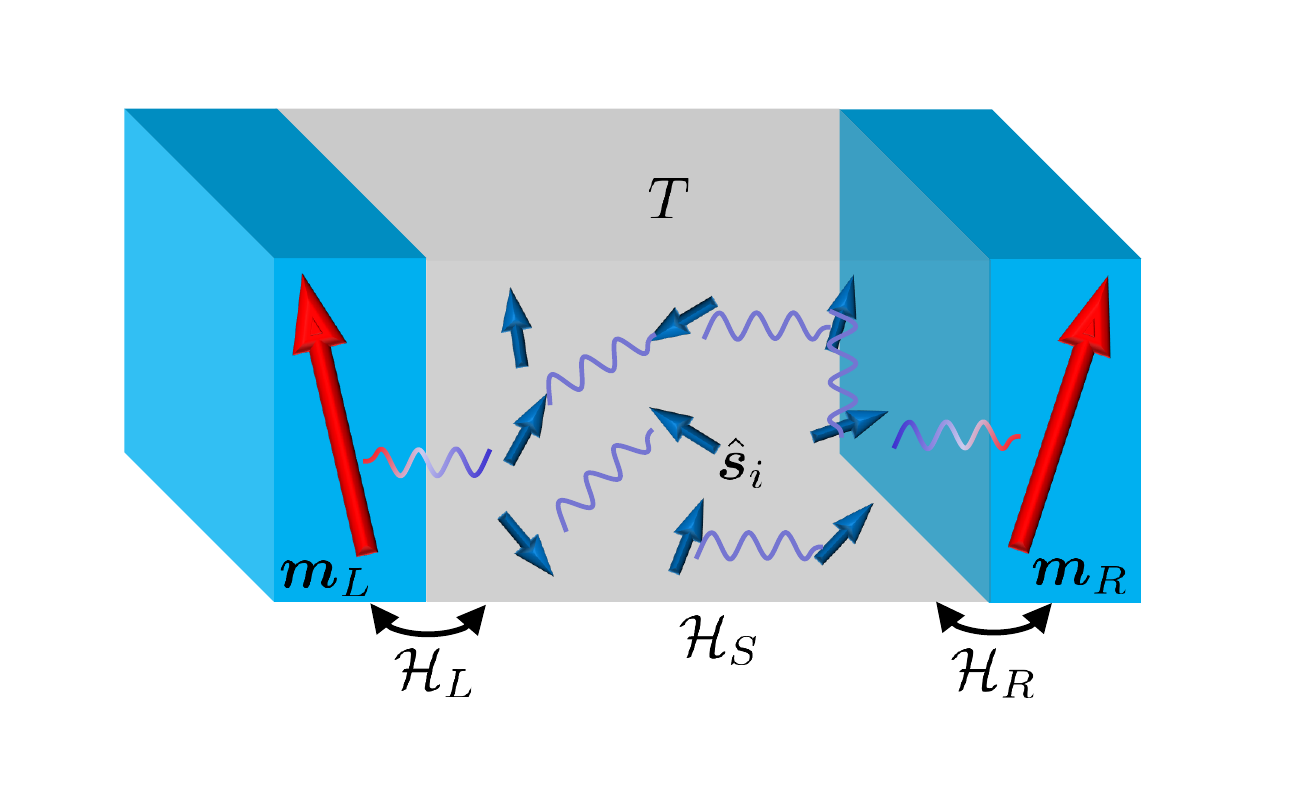} 
	\caption{Illustration of the set-up consisting of two single-domain macroscopic magnets, with magnetization direction $\bM_L$ and $\bM_R$ for left and right ferromagnet, respectively, that are coupled to an intermediate system with spins $\hat \bs_i$. Here, $i$ labels the spins of the intermediate system that is in thermal equilibrium at temperature $T$.}
	\label{fig:setup}
\end{figure}

The magnetic interactions described above arise from the minimization of energy, usually the energy of the electrons that mediate the interactions \cite{RevModPhys.95.035004}. In this Letter, we propose interactions between two magnetic layers that arise from maximization of the entropy, which is now what mediates the interactions. Entropic interactions are common in biophysical and soft-matter systems. Well-known examples include the entropic spring that results from the configurational entropy of polymers \cite{rubinstein2003polymer} and depletion interactions in colloidal systems \cite{asakura1954,Lekkerkerker1992, dijkstra1999phase, Lekkerkerker2011ColloidsAT}. Examples of entropic interactions in other physical systems include engineered entropic pressure gradients in optical cavities \cite{doi:10.1126/sciadv.ade3591}. It was even suggested that gravity is fundamentally an entropic force \cite{Verlinde2011}. 

The entropic magnetic interlayer coupling that we propose here may eventually become useful for applications, for example because --- contrary to energy-dominated interactions --- the resulting torques dominate over some range of temperatures and increase in strength over this range. This may be beneficial for thermal stability, which is an obstacle for many applications.

{\it Model and general theory.} --- The set-up we consider in this Letter is illustrated in Fig.~\ref{fig:setup}. Two macroscopic single-domain magnets are connected on the left and right of an intermediate microscopic spin system. The magnetic moments of the single-domain magnets are denoted by $\bM_L$ and $\bM_R$ for the left and right magnet, respectively. The spins of the intermediate system are denoted by $\hat \bs_i$, where $i$ labels the $N$ sites of the spin system. The set-up may be one, two, or three-dimensional. We consider $\bM_L$ and $\bM_{R}$ to be classical variables, which is appropriate in the case that their saturation magnetization is sufficiently large. This assumption is, however, not necessary but merely facilitates the singling out of the magnetic interlayer coupling. The spins of the intermediate system may be treated either classically or quantum-mechanically, depending on the choice of material and model. These latter spins are assumed to be in thermal equilibrium at temperature $T$ (for example due to coupling to the lattice vibrations of the underlying lattice), such that the intermediate spin system explores its microstates. This temperature is assumed to be much smaller than the Curie temperature of both macroscopic magnets.

The coupling between the left ferromagnet and the spins of the intermediate system is described by the Hamiltonian $\hat {\mathcal H}_L [\bM_L, \hat \bs_i]$, whereas the coupling to the right magnet is described by $\hat {\mathcal H}_R [\bM_R, \hat \bs_i]$. The Hamiltonian for the intermediate spin system is denoted by $\hat {\mathcal H}_S [\hat \bs_i]$. Since we are interested in the coupling between the left and right macroscopic magnets that is mediated by the spin system, we do not, for the moment, consider the energetics of the macroscopic magnets due to their intrinsic anisotropy or interactions with an external field. 

Keeping $\bM_L$ and $\bM_{R}$ fixed, the partition function is given by 
\begin{eqnarray}
 && Z (\bM_L, \bM_R, T, N) \nonumber \\ && = {\rm Tr} \left[ e^{-\beta \left(\hat {\mathcal H}_L [\bM_L, \hat \bs_i] +\hat {\mathcal H}_R [\bM_R, \hat \bs_i]+{\mathcal H}_S [\hat \bs_i]\right)}\right]~,
\end{eqnarray}
where $\beta=1/k_B T$ is the inverse thermal energy, $k_B$ is Boltzmann's constant, and the trace is taken over the Hilbert space of the intermediate spin system. For the case of a classical spin system, this trace will be an integration over classical spin vectors. We now use that the free energy $F (\bM_L, \bM_R, T, N) =-k_B T \ln Z  (\bM_L, \bM_R, T, N)= E (\bM_L, \bM_R, T, N) -  T S (\bM_L, \bM_R, T, N) $, in terms of the internal energy $E (\bM_L, \bM_R, T, N)$ and entropy $S (\bM_L, \bM_R, T, N)$. This free energy describes effective magnetic interactions between the macroscopic magnetizations $\bM_L$ and $\bM_R$ that are mediated by the spin system at temperature $T$. For example, a term $F (\bM_L, \bM_R, T, N) \propto \bM_L \cdot \bM{_R}$ would imply an effective exchange interaction between the two macroscopic magnetizations. In the remainder of this Letter we focus on the situation that this effective magnetic interlayer coupling is dominated by maximization of the entropy $S (\bM_L, \bM_R, T, N)$ rather than minimization of the energy $E (\bM_L, \bM_R, T, N)$. We refer to the entropy-dominated case as entropic magnetic interlayer coupling.

Next, we discuss an example of a spin system that mediates entropic magnetic interactions. In the Supplementary Information \cite{suppinfo}, we discuss an additional example. These examples share the feature that geometric frustration gives rise to configurational spin entropy \cite{lacroix2011introduction} that depends on the coupling of the spins to the macroscopic magnets and mediates the entropic interlayer coupling. 

\begin{figure} 
	\includegraphics[width=9cm]{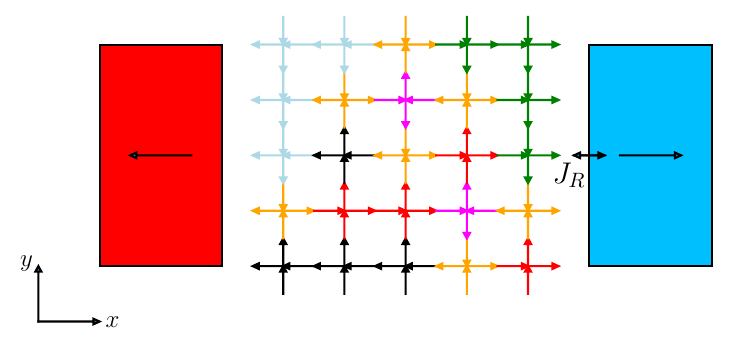} 
	\caption{Illustration of the set-up in which the magnetic interlayer coupling is mediated by square spin ice. The left and right magnet are strongly exchange coupled to the intermediate spins so that the direction of the spins adjacent to the magnets is fully determined by the direction of the magnet. The six different vertices of the square spin ice are indicated with different colours.}
	\label{fig:sixvertex}
\end{figure}

{\it Example: Square spin ice.} --- As a concrete model for the intermediate spin system we consider square spin ice, as modelled by the six-vertex model (see. Fig.~\ref{fig:sixvertex}). This model consists of classical spins on and pointing along the bonds of a square lattice, whose dimensions are $N_x \times N_y$ \cite{baxter1982exactly}. The spins are subject to the ice rules which dictate that each vertex has two spins pointing in and two pointing out. This yields six allowed vertex configurations. This model is a simple model for spin-ice materials \cite{spinicereview}, as well as for artificial spin ice metamaterials \cite{Skjarvo2020}, both of which have attracted a great deal of attention. The free energy of the six-vertex model is known to depend on the boundary conditions, even in the thermodynamic limit \cite{VKorepin2000}. Hence, we expect this model to mediate an entropic interaction.

\begin{figure} 
	\begin{center}
	\includegraphics[width=10cm]{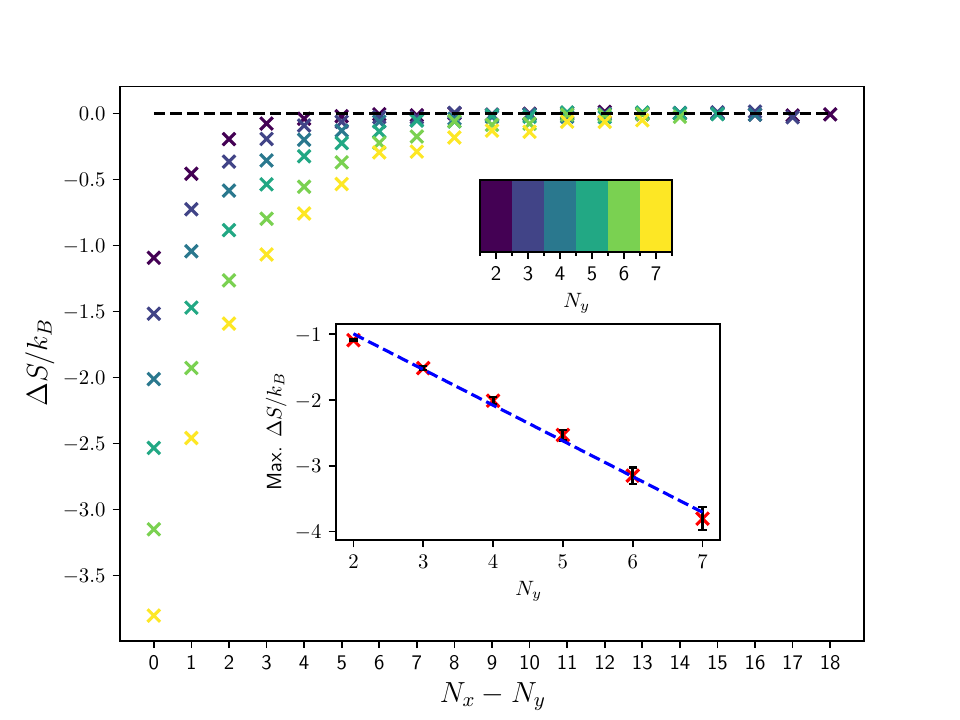} 
	\caption{Entropy difference between parallel and antiparallel states of two magnets whose interaction is mediated by square spin ice. The horizontal axis corresponds to the number of vertices in the square spin ice in the horizontal direction.}
	\label{fig:entropydifferencesixvertex}
\end{center}
\end{figure}

To facilitate the computation of the entropic magnetic interaction we assume that the coupling between the left macroscopic magnet and the spins on the leftmost boundary of the square ice is so strong that it pins the spins in the direction of the left macroscopic magnet. For the coupling between the right spins and the right macroscopic magnet, we assume a coupling ${\mathcal H} [{\bm m}_R, {\bm s}_i] = -J_R \sum_i' {\bm s}_i \cdot {\bm m}_R$, where the sum is over the spins on the right edge of the intermediate spin system (see Fig.~\ref{fig:sixvertex}). We take the left magnet to be pointing to the left, and allow the right magnet to point to left or right. For this model, the energy difference between parallel ($J_R \to -\infty$) and antiparallel ($J_R \to +\infty$) configurations of the macrosopic magnets is zero in the limit that the coupling of the spins to these magnets is infinitely strong. Hence, the free-energy difference only has a contribution from the entropy difference, which is found through thermodynamic integration to be \cite{suppinfo}
\begin{equation}
	\frac{\Delta S}{k_B} =\beta\int_{-\infty}^\infty \left\langle \sum_i' s_i \right\rangle dJ_R~,
\end{equation}
where the sum over $s_i =+1~(-1)$ corresponds to the spins on the right-most boundary pointing to the right (left). The thermal average $\langle \cdots \rangle$ in the above is evaluated using Monte Carlo simulations. Specifically, we use the short-loop algorithm \cite{newmanb99} and discard any move that affects the left-most spins. In this way, they are effectively pinned by the left magnet. We use open boundary conditions for the spins at the top and bottom edge, corresponding to the experimental situation of artificial spin ice. \\
In Fig.~\ref{fig:entropydifferencesixvertex}, we show the entropy difference between anti-parallel and parallel configurations of the macroscopic spins, as a function of the size of the intermediate system, which corresponds to the number $N_x$ of vertices between the left and right magnets in the horizontal direction. The curves corresponding to different values of $N_y$ are offset by $N_y$, so that all the curves start at the origin. \\
The results clearly show an entropy difference --- and therefore a entropic magnetic interlayer coupling --- that decays with the size of the system. The precise functional nature of this decay is difficult to determine from the simulations, although it is certainly slower than exponential. All the entropy differences are negative, implying that a parallel configuration of the two macroscopic magnets has a higher entropy. \\
The six-vertex model is a critical model with infinite correlation length \cite{baxter1982exactly}. Hence, the slower-than-exponential decay is to be expected. We remark that since, with such small systems, we are not in the thermodynamic limit, comparison to analytical results for the free energy obtained in this limit remains challenging. \\
In the inset of Fig ~\ref{fig:entropydifferencesixvertex}, we show the maximum entropy difference $\Delta S / k_B$ as a function of the vertical size of the system, corresponding to the number of vertices on the interfaces that feel the coupling to the left and right magnets. This maximum always corresponds to the case of a square system ($N_x = N_y$). The maximum entropy difference increases with $N_y$, and appears to increase in a nearly-linear fashion. \\
We note that because of the critical nature  of square spin ice, the decay of the entropic magnetic interlayer coupling with distance is similar to the Casimir and critical Casimir interactions  \cite{RevModPhys.71.1233,RevModPhys.90.045001,PhysRevLett.114.038301}. In these interactions, an algebraically decaying force arises from the confinement of quantum fluctuations of the electromagnetic field and from critical density or composition fluctuations of the solvent, respectively. With the exception of Ref.~\cite{PhysRevLett.88.240401}, which considers a magnetic interlayer coupling due to vacuum fluctuations, Casimir interactions in spin systems \cite{PhysRevB.92.214409} have --- though the set-up we consider is reminiscent of the set-up for the wetting transition in spin systems, see e.g. Ref.~\cite{PhysRevLett.116.046101} --- to the best of our knowledge not been considered as a mediator of magnetic interlayer coupling. Moreover, the entropic magnetic interlayer coupling that we propose here is general and does not require the mediating spin system to be critical. We also note that while the effects of an interface on the magnetic properties of spin ice thin films have been extensively studied, for example in Refs. \cite{Bovo2014,Bovo2019, Jaudert2017, She2017, Lantagne2018} , the effect of a magnetic interface on (artificial) spin ice remains an open question.


Having given an example of an entropy-dominated magnetic interlayer coupling, we now turn to a general discussion of the broader consequences of such couplings for the dynamics of the left and right macroscopic magnets.

{\it Entropic torques.} --- In the case that $N$ is sufficiently large such that the dynamics of the individual microscopic spins add up to a mean field, of which the fluctuations can in the first instance be ignored, and in the case that the relaxation time of the spin system is much shorter than the precession period of both macroscopic magnets, the intermediate spin system continuously explores many microstates while the macroscopic magnets undergo their slow dynamics. Then, the  magnetic interactions yield a torque $\gamma_{L(R)}  \bM_{L (R)} \times \partial F/\partial (M_{s,L(R)}\bM_{L (R)})$, which contains an entropic torque $-\gamma_{L(R)} T \bM_{L (R)} \times \partial S/\partial (M_{s,L(R)}\bM_{L (R)})$ on the left (right) ferromagnet. This entropic torque should be added to the Landau-Lifshitz-Gilbert equation that describes the classical magnetization dynamics. Here, $M_{s, L (R)}$ is the saturation magnetization of the left (right) ferromagnet, and $\gamma_{L(R)}$ is its gyromagnetic ratio. The
torques and effective fields due to the entropic interactions and in principle be very large. Taking the result for square spin ice in Fig. \ref{fig:entropydifferencesixvertex}, we find that $\Delta S/k_B$ can be of order one. From this, one estimates an effective field of order $T\Delta S/\mu_B \sim k_B T/\mu_B$.

{\it Mutual information.} --- We now discuss the situation of finite $N$ in more detail. While a mean-field-like torque as discussed above cannot be defined in this case, we will argue that the system is characterized in a useful way by the mutual information between the two macroscopic magnets. We assume that the left and right macroscopic magnets and their coupling to the intermediate spin system are identical --- a situation we refer to as the symmetric case.  For simplicity, we assume in what follows that the anisotropy of the macroscopic magnets is such that their low-energy states favour alignment with the $z$-axis. We are then concerned with four possible states of the macroscopic magnets whose entropy we denote as $S_{\alpha\alpha'} (T,N)$, with $\alpha, \alpha' \in \{\uparrow, \downarrow \}$. To be more specific, we have $S_{\uparrow\uparrow} (T,N)=S (\bM_L=\hat {\bz}, \bM_R=\hat \bz, T, N)$, $S_{\downarrow\uparrow} (T,N)=S (\bM_L=-\hat {\bz}, \bM_R=\hat \bz, T, N)$, $S_{\uparrow\downarrow} (T,N)=S (\bM_L=\hat {\bz}, \bM_R=-\hat \bz, T, N)$, and $S_{\downarrow\downarrow} (T,N)=S (\bM_L=-\hat {\bz}, \bM_R=-\hat \bz, T, N)$, with $\hat \bz$ the unit vector in the $z$-direction. In the symmetric situation, we have for the antiparallel case that $S_{\uparrow\downarrow} (T,N)=S_{\downarrow\uparrow} (T,N) \equiv S_A (T,N)$, and $S_{\uparrow\uparrow} (T,N)=S_{\downarrow\downarrow} (T,N) \equiv S_P (T,N)$ for the parallel case, where (anti)parallel refers to (mis)alignment of the macroscopic magnets. For future reference, we introduce the energies $E_{\alpha\alpha'} (T,N)$ for the energy of the system when the left (right) magnet is in state $\alpha (\alpha') \in \{\uparrow, \downarrow\}$, and we introduce  $E_P (T,N)$ and $E_A (T,N)$ as the energies of the parallel and antiparallel state.

\begin{figure} 
	\includegraphics[width=8.5cm]{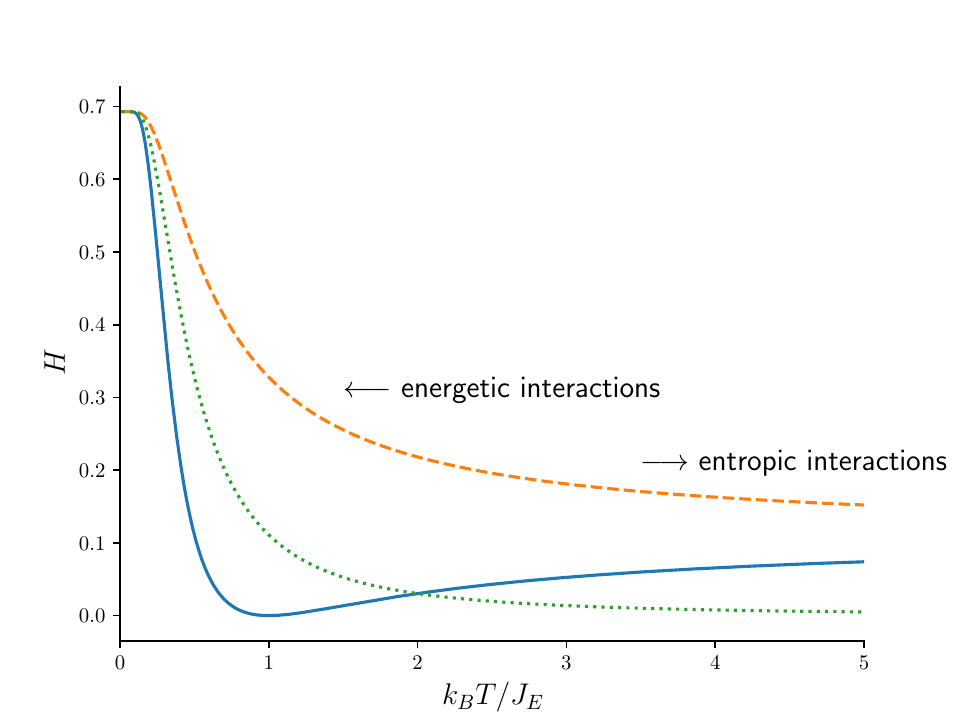} 
	\caption{Mutual information of two magnets coupled by energetic and entropic interactions. The solid and dashed curve correspond to competing and cooperating energetic and entropic interactions, respectively. The dotted curve corresponds to purely energetic interactions. For small temperature the energetic interactions dominate the mutual information, whereas the entropic interactions lead to a non-zero mutual information even for high temperatures.}
	\label{fig:mutualinfo}
\end{figure}

Let us now discuss what it means physically when $S_P (T,N) \neq S_A (T,N)$. In this discussion, it is important to mention that the systems we are interested in, such as the artificial spin ices discussed before, are not necessarily in the thermodynamic limit in which $N \to \infty$. We assume that the anisotropy in the macroscopic magnets that pins their magnetization to ``up" or ``down" is not much larger than the thermal energy. Then, the left and right magnet will undergo thermally-activated dynamics. While the system undergoes this dynamics, the difference in odds of finding the macroscopic magnets in the parallel state are given by  $e^{(S_P-S_A)/k_B-\beta(E_P-E_A)}$. The difference between the energy-dominated ($S_A=S_P$) and entropy-dominated ($E_A=E_P$) case is their temperature dependence. For the energy-dominated case, the system will always go to the state with lowest energy for $T \to 0$ even when $N$ remains finite, while for high temperature the parallel and antiparallel states are equally likely. For the entropy-dominated case, the state of the system is probabilistic even when $T$ becomes small (the limit that $T\to 0$ should be considered by explicitly taking into account the spin dynamics, as the spin system likely no longer explores its microstates when  $T\to 0$), and remains probabilistic when $T$ becomes large. For large entropy differences between parallel and antiparallel states, however, the difference in odds of finding the system in either parallel or antiparallel state will be very large even for high temperatures.

Though the configuration of two macroscopic magnets that are coupled by entropic interactions is probabilistic, especially when $N$ is small, the set-up is still useful to transmit information. This information transmission is a ``noisy" communication channel, in the sense introduced by Shannon \cite{shannon}. Here, it is useful to consider one of the macroscopic magnets, say the left one, as the input, and the other as the output of the channel. For example, one may ask the question: if the left magnet is fixed to point up, what is the probability for the right magnet to point up as well? In other words, how much information does a measurement of the output (the state of the right magnet) convey about the input (the state of the left magnet)? The quality of such a communication channel is characterized by the mutual information
\begin{equation}
\label{eq:mutinfo}
H=\sum_{\alpha,\alpha'} P_{\alpha\alpha'}
\ln \left( \frac{P_{\alpha\alpha'}}
	{P_{L,\alpha} P_{R,\alpha'}} \right),
\end{equation}
that quantifies how much a measurement of the output reduces the uncertainty on the state of the input. Here, $P_{\alpha\alpha'}$ is the probability of finding the left magnet in state $\alpha$ and the right magnet in state $\alpha'$. The probability of finding the left (right) magnet in state $\alpha$ is denoted by $P_{L(R), \alpha}$. For the symmetric situation under consideration we have that $P_{L(R), \alpha}=1/2$ as both "up" and "down" states are equally likely. When the two magnets are not correlated, we have that $P_{\alpha\alpha'} = P_{L,\alpha} P_{R,\alpha'}$ so that $H=0$. Using that $P_{\alpha\alpha'} \propto e^{ -\beta [E_{\alpha\alpha'} (T,N)-TS_{\alpha\alpha'} (T,N)]}$, one finds that $H$ goes to zero for high temperatures in the case of purely energetic interactions ($S_A=S_P$). For small temperatures and energetic coupling, the mutual information reaches its maximum value of $\ln 2$ as the magnetic configuration is then in the low-energy state and measurement of the state of one magnet fully determines the the state of the other. For entropic interactions ($E_A=E_P$), however, the mutual information is only weakly dependent on temperature $T$ through the dependence of the entropy on temperature. For higher temperatures, the mutual information does not go to zero as in the case of purely energetic interactions, but rather saturates to the value $H=\sum_{\alpha\in \{+,-\} } [1+e^{\alpha \Delta S/k_B}]^{-1} \ln \{2/[1+e^{\alpha \Delta S/k_B}] \}$, with $\Delta S = S_P-S_A$. For large $\Delta S$ we therefore find that $H$ obtains its maximum value of $\ln 2$ even at high temperatures, which is not possible with purely energetic magnetic interactions. This feature is promising for applications that need to operate in an environment with thermal fluctuations.

In Fig.~\ref{fig:mutualinfo} we plot the mutual information for $E_{\alpha\alpha'}=-J_E\alpha\alpha'$ and $S_{\alpha\alpha'}=-J_S\alpha\alpha'$, where $\alpha, \alpha'$ take the value $+1$ ($-1$) when the index $\alpha,\alpha'$ is $\uparrow$ ($\downarrow$). For the solid and dashed curves we take $|J_E|=|J_S|$. The solid curve is for competing energetic and entropic interactions ($0<J_E=-J_S$) which leads to a minimum in the mutual information where the two magnets are effectively uncoupled. The dashed curve is for $0<J_E=J_S$. The dotted curve is the result for purely energetic interactions $(J_E \neq 0, J_S=0)$. From this figure, it is clear that the mutual information remains finite at high temperatures due to entropic interactions.

{\it Discussion and conclusion.} --- We have provided a general theory of entropic magnetic interlayer coupling and discussed its consequences both for large systems and small systems. In the former, the entropic interactions give rise to entropic torques on the magnetization direction. For small systems, the consequences of entropic interactions were discussed in terms of the mutual information between the coupled magnets.

Future work could consider dynamics of the intermediate system and the magnets that are coupled, as well as interactions with a spin current at an interface between the entropy-dominated spin system and a heavy metal \cite{han2020}. Motivated by the examples that we provided, an interesting direction is to explore frustrated magnets \cite{lacroix2011introduction} as systems that mediate entropy-dominated coupling between magnets. Experimentally, it would be interesting to directly measure the mutual information between two coupled small magnets, for example by using the techniques put forward in Ref.~\cite{adhikari2023magnetization}. Some aspects of an artificial frustrated system which is coupled to two larger magnets have been studied experimentally \cite{khajetoorians2011realizing}, so it would be interesting to build upon this work.

The entropic magnetic interlayer coupling becomes stronger upon increasing the temperature. For temperatures that are much larger than the internal interaction-energy scale of the mediating system, however, the interlayer coupling will go to zero. For example, the interlayer coupling that is mediated by square spin ice will diminish for thermal energies that are so large that the spins become decoupled, which allows for violation of the ice rules. In general, therefore, frustrated magnets with large Curie-Weiss temperature may be attractive for engineering magnetic layer interactions that increase over a large temperature window.

\bibliography{biblio}

\section{Supplementary information}
\subsection{Analytic partition function}
\begin{figure}[h!]
	\centering
	\includegraphics[width=0.5\textwidth]{{ 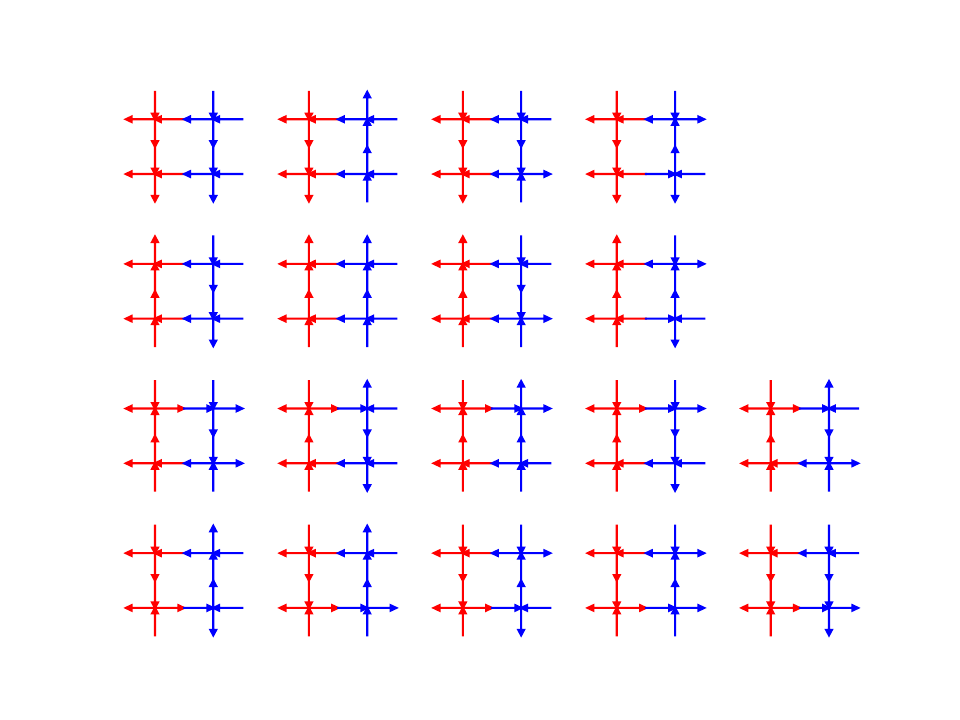 }}
	\caption{All possible configurations of a 2x2 system of vertices, subject to the constraint that the leftmost spins are fixed to point to the left.}
	\label{fig:all_analytic_2x2_configs}
\end{figure}
\begin{figure}
	\centering
	\includegraphics[width=0.5\textwidth]{{ 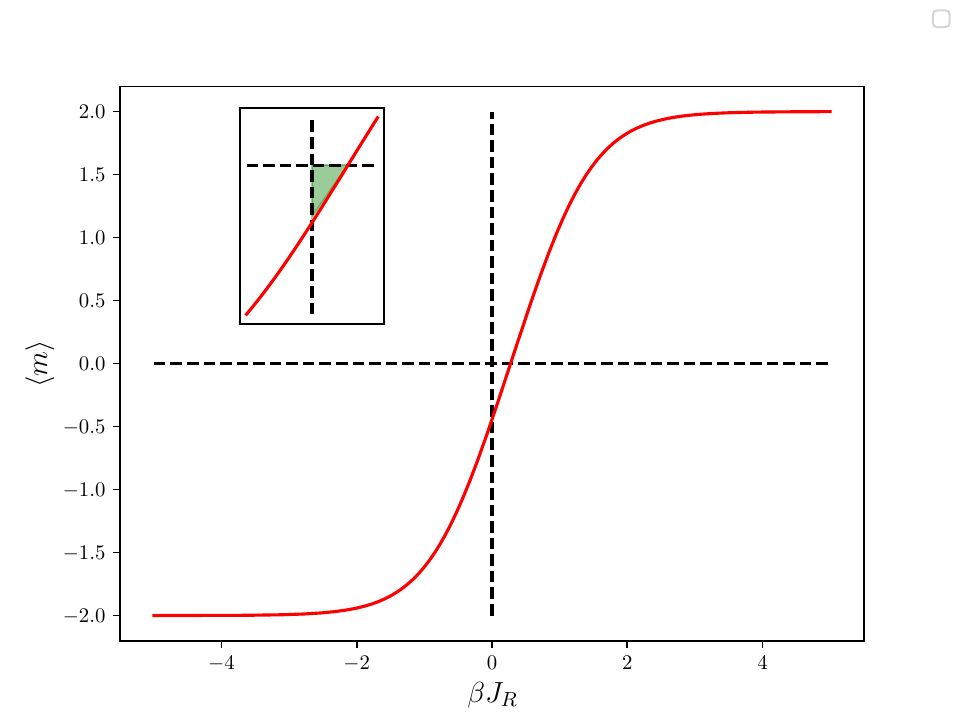 }}
	\caption{Average magnetisation as a function of dimensionless coupling $\beta J_R$ for the 2x2 analytic case. The green area in the inset plot highlights the asymmetry in the curve, which is the predominant contribution to the entropy difference.}
	\label{fig:analytic2x2_magnetisation}
\end{figure}

For sufficiently small systems, it is feasible to compute the analytical partition function. It is instructive to demonstrate how this works so that the entropic contribution to the free energy is more readily seen. We therefore consider a system of $2\times2$ vertices. \\
The left magnet is assumed to point to the left, and the coupling of the leftmost vertices thereto is sufficiently large such that the leftmost spins are also fixed to point to the left, which significantly restricts the possible states that the leftmost vertices can assume. \\
 In our simulations, we adopt the convention that spins pointing to the right have positive magnetisation $+1$, whereas those which point to the left have negative magnetisation, $-1$. We therefore use the same convention here. We illustrate all the possible configurations of the system subject to these constraints in Fig. \ref{fig:all_analytic_2x2_configs} , and compute the energies and magnetisations of each configuration according to the above convention using the Hamiltonian 
 \begin{equation}
 	H = -J_R \sum_{i\in B_R}\sigma_{iR}m_R,
 \end{equation}
 where $J_R > 0 $, $\left\{\sigma_{Ri}\right\}$ are the spins on the right interface, and $m_R = +1$ is the state of the right magnet, which is fixed to point to the right. Note that varying the sign of $J_R$ from positive to negative models the possibility that the right magnet is in fact parallel with the left one. \\
 We find that the partition function is
 \begin{equation}
 	Z = 10 + 6 e^{-2\beta J_R} + 2e^{2\beta J_R},
 \end{equation}
while the expectation value of the magnetisation on the right interface is found to be
\begin{equation}
	\langle m \rangle = \frac{-6e^{-2\beta J_R} + 2e^{2\beta J_R}}{5 + 3e^{-2\beta J_R} + e^{2\beta J_R}},
\end{equation}
which is plotted in Fig. \ref{fig:analytic2x2_magnetisation}.
 \\
Using Eq. (2) from the main text, we find the entropy difference by integrating $\langle m\rangle$ from $-\delta \equiv -\beta J_R$ to $+\delta$, obtaining
\begin{align}
	\int_{-\delta}^{+\delta} \langle m \rangle d(\beta J_R) &= \left[\ln\left(5 + 3e^{-2\beta J_R} + e^{2\beta J_R}\right)\right]_{\beta J_R = -\delta}^{\beta J_R = +\delta} \nonumber \\
	&= \ln\left(\frac{5 + 3e^{-2\delta} + e^{2\delta}}{5 + 3e^{2\delta}  + e^{-2\delta}}\right).
\end{align}
\begin{figure}
	\centering
	\includegraphics[width=0.5\textwidth]{{ 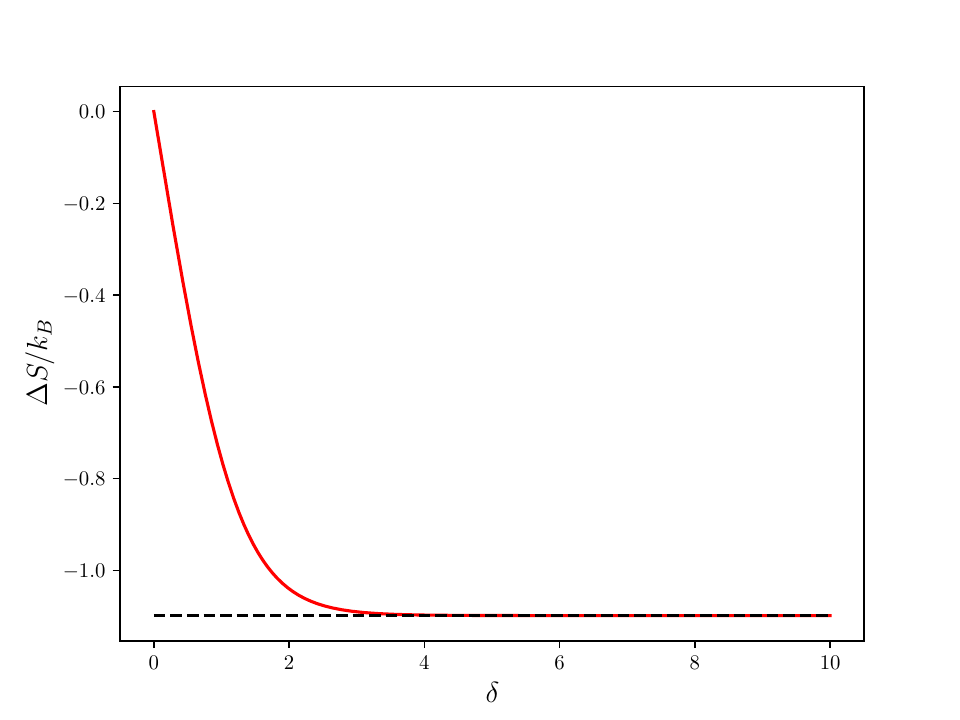 }}
	\caption{Entropy difference as a function of maximum coupling $\delta$, indicating that the entropy difference approaches a constant value in the infinite-coupling limit.}
	\label{fig:mag_integrated_2x2}
\end{figure}
In the limit that $\delta \to\infty$, this integrand approaches $-\ln 3$, as shown in Fig. \ref{fig:mag_integrated_2x2}. This indicates that the entropy difference between the antiparallel and parallel configurations of the left and right magnets is negative. Therefore, the parallel configuration has a higher entropy, which can also be seen directly from the partition function.
\subsection{Thermodynamic integration}
Our starting point is the Hamiltonian which describes the coupling of the spins on the right interface to the large magnet. We choose this to be
\begin{equation}
	H_{\text{int}} = -J_R\sum_{i\in B_R} \sigma_i,
\end{equation}
where the convention is that $\sigma_i = \pm 1$ for right (left) pointing interface spins, such that for $J_R  > 0$  the right magnet is pointing to the right, and the two macroscopic magnets are in an anti-parallel configuration, whereas for $J_R < 0$ the right magnet is pointing to the left, and the two macroscopic magnets are in a parallel configuration. \\ 
We choose this convention because, in our simulations, we initialise the left interface vertices in a state where the leftmost spins are pointing to the left. 
We next construct the partition function, which is given as a sum over possible configurations of interface spins of the Boltzmann weights of each configuration,
\begin{equation}
	Z = \sum_{\left\{\sigma_i\right\}} e^{-\beta H_{\text{int}}[\left\{\sigma_i\right\}]} = \sum_{\left\{\sigma_i\right\}} e^{\beta J_R \sum_{i\in B_R} \sigma_i}.
\end{equation}
Rearranging the expression for the free energy $F = E - TS = -\frac{1}{\beta}\ln Z$ to find the (dimensionless) entropy yields
\begin{equation}
	\frac{S}{\kB} = \beta E + \ln Z,
\end{equation}
which, upon differentiation with respect to $J_R$, becomes
\begin{equation}
	\frac{1}{\kB}\frac{\partial S}{\partial J_R} = \beta\frac{\partial E}{\partial J_R} + \frac{1}{Z}\frac{\partial Z}{\partial J_R}.
\end{equation}
The second term is
\begin{align}
	\frac{1}{Z}\frac{\partial Z}{\partial J_R} &= \frac{1}{Z}\frac{\partial}{\partial J_R}\left(\sum_{\left\{\sigma_i\right\}}e^{\beta J_R\sum_{i\in B_R}\sigma_i}\right) \nonumber \\
	&= \frac{\beta}{Z}\sum_{\left\{\sigma_i\right\}}\left(\sum_{i\in B_R}\sigma_i\right)e^{\beta J_R\sum_{i\in B_R}\sigma_i} \nonumber \\
	&= \frac{\beta}{Z}\sum_{\left\{\sigma_i\right\}}\left(\sum_{i\in B_R}\sigma_i\right)e^{\beta H(\left\{\sigma_i\right\})} \nonumber \\
	&= \beta\Braket{\sum_{i\in B_R}\sigma_i}
\end{align}
where the brackets $\langle\cdots\rangle$ denote a thermodynamic average, which we can equate with an ensemble average sampled from Monte Carlo simulations. The rate of entropy change with respect to $J_R$ is then
\begin{equation}
	\frac{1}{\kB}\frac{\partial S}{\partial J_R} =  \beta\frac{\partial E}{\partial J_R} + \beta\Braket{\sum_{i\in B_R}\sigma_i}.
\end{equation}
An anti-parallel configuration of the two magnets corresponds to an infinite positive interface coupling $J_R \to +\infty$, whereas a parallel configuration corresponds to an infinite negative interface coupling $J_R \to -\infty$. We therefore integrate the expression for $\partial S/\partial J_R$ over this range to obtain the entropy difference between such configurations as
\begin{equation}
	\frac{\Delta S}{\kB} = \beta\left(\Delta E + \int_{-\infty}^\infty\Braket{\sum_{i\in B_R}\sigma_i}d J_R\right).
\end{equation}
There is one more step to arrive at the equation quoted in the main text. To do this, we note that for large enough positive (negative) coupling $J_R$, the interface spins all point to the right (left), and the net magnetisation on the right interface is $\pm N_y$, where $N_y$ is the number of spins at the interface. Therefore, the interface energy for some finite positive coupling $+J_{R,\text{max}}$ is $E_+ = (-J_{R,\text{max}})(N_y) = -J_{R,\text{max}}N_y$, whereas for finite negative coupling $-J_{R,\text{max}}$ it is $E_- = -(-J_{R,\text{max}})(-N_y)  = -J_{R,\text{max}} N_y$. So, $E_+ = E_-$ and the energy difference $\Delta E = 0$. Hence,
\begin{equation}
	\frac{\Delta S}{\kB} = \beta\int_{-\infty}^\infty\Braket{\sum_{i\in B_R}'\sigma_i}dJ_R,
\end{equation}
which is the quoted result in Eq. (2).

\subsection{Paramagnetic spins}
\begin{figure} 
	\includegraphics[width=8cm]{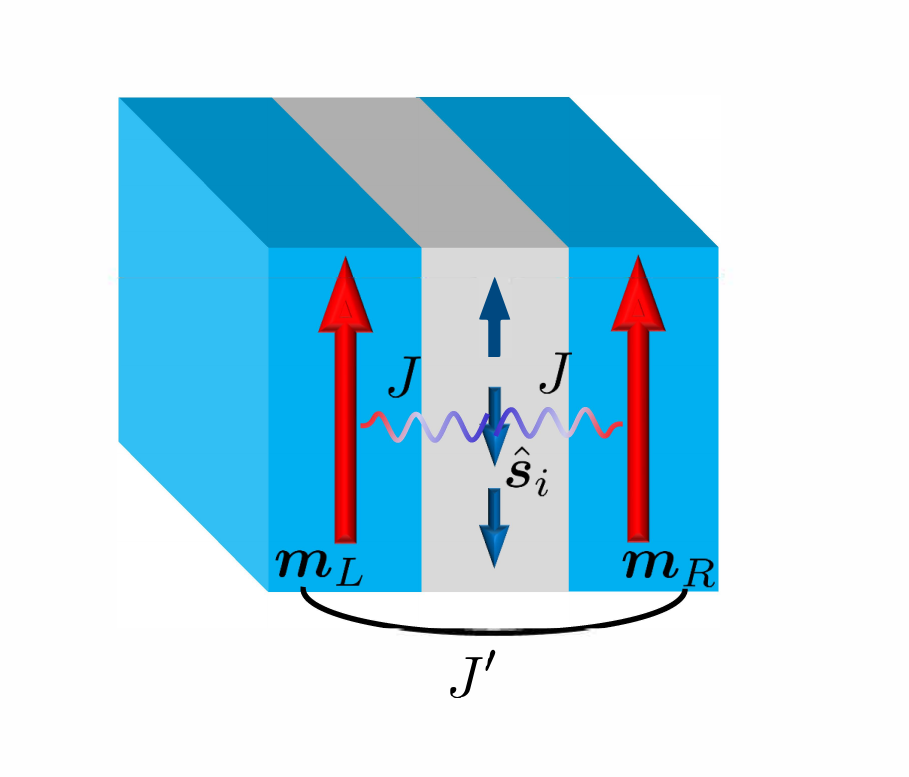} 
	\caption{Two macroscopic magnets that are coupled directly via an antiferromagnet exchange interaction with strength $J'$ and are coupled antiferromagnetically (with strength $J$) to $N$ paramagnetic spins at temperature $T$.}
	\label{fig:paramagnetic}
\end{figure}

As another example of a system that mediates an entropy-dominated magnetic interlayer coupling, we consider the simple model that is illustrated in Fig.~\ref{fig:paramagnetic}. Here, the coupling between the macroscopic magnets is mediated via paramagnetic spins $\hat \bs_i$ to which both macroscopic magnets couple antiferromagnetically, with exchange constant $J>0$. We consider the paramagnetic spins to be Ising spins, such that $\hat \bs_i \equiv s_i = \pm 1  $. We focus on the situation that the macroscopic magnets are pointing in the positive or negative $z$-direction, so we have that $\bM_{L/R} = m_{L/R} \hat z$, with $m_{L/R} = \pm 1$. In addition to the magnetic coupling that is mediated via the paramagnetic spins, we also consider a direct antiferromagnetic coupling between the two magnetic layers with exchange constant $J'>0$. While we are not aware of a specific physical realization of this model, it is not unlikely that it could be realized by stacking thin layers of magnetic and paramagnetic materials. In addition, it could also be engineered synthetically, by coupling small single-domain nanomagnets with uniaxial anisotropy. 

The Hamiltonian for the system is
\begin{equation}
\label{eq:hamparamagnet}
 {\mathcal H} = J' m_L m_R + J (m_L+m_R) \sum_{i=1}^N s_i~.
\end{equation}
At zero temperature, the energy of the parallel configuration of the macroscopic magnets is $E_P=J'-2JN$, whereas the energy of the antiparallel configuration is $E_A=-J'$. At zero temperature, the magnetic interlayer interaction is therefore effectively ferromagnetic when $J'<2JN$. We assume in the following that $J'$, $J$, and $N$ are chosen such that this is the case. 

For an antiparallel alignment of the macrosopic magnets, the paramagnetic spins contribute $\ln 2$ per spin to the entropy independent of the temperature. This is because the antiferromagnetic coupling $J$ leads to frustration of the paramagnetic spins. That is, when the macroscopic magnets are antiparallel, the microscopic Ising spins can point up or down with equal energy. As a result, the entropy for the antiferromagnetic alignment of the macroscopic magnets is larger. Therefore, we expect a transition upon increasing the temperature from an energy-dominated ferromagnetic interlayer coupling to an entropy-dominated antiferromagnetic interlayer coupling. 
\begin{figure}[ht!]
	\includegraphics[width=8.5cm]{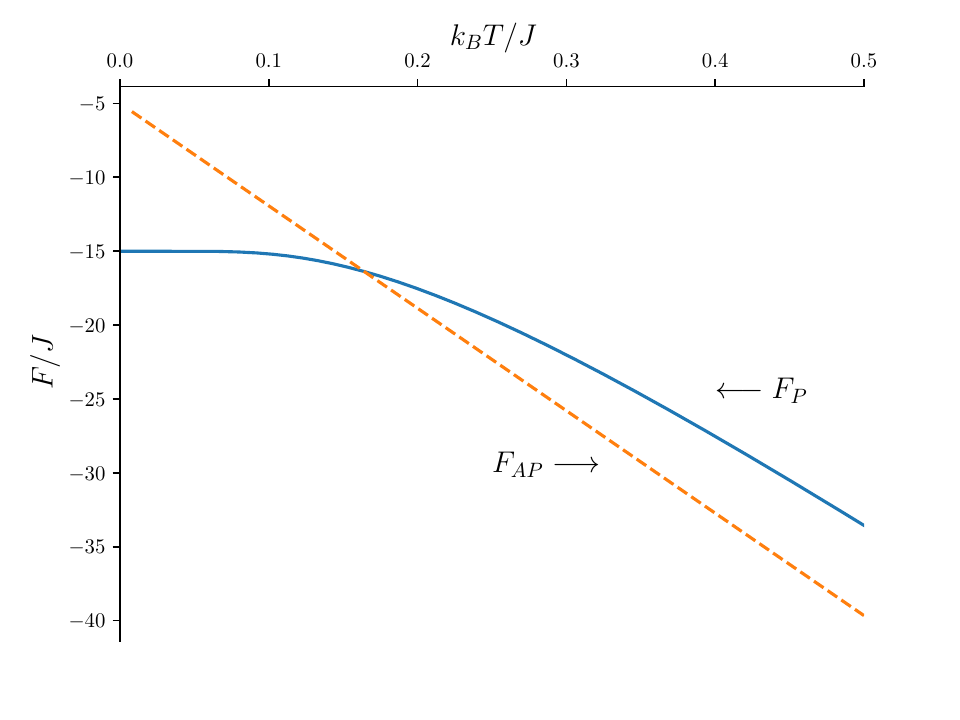} 
	\caption{Free energy as a function of temperature for parallel configuration of the macroscopic magnets (solid curve) and antiparallel configuration (dashed curve). The parameters taken are $J'/J=50$ and $N=100$. }
	\label{fig:deltafparamagnet}
\end{figure}
The free energy of the Ising spins for an antiparallel alignment of the macroscopic magnets is $F_A=-J'-k_BTN\ln 2$. For small temperatures $k_B T \ll J$, the free energy of the parallel alignment is $F_P\simeq J'-2JN$. As a result, we expect the transition from effectively ferromagnetic to effectively antiferromagnetic interlayer coupling to occur at the temperature $k_B T^* \simeq (2JN-J')/N \ln 2$. In Fig.~\ref{fig:deltafparamagnet} we plot the free energies of the parallel and antiparallel configurations of the macroscopic magnets. These free energies are computed from the partition function corresponding to the Hamiltonian in Eq.~(\ref{eq:hamparamagnet}), as described in the main text. As expected, the free energy for the parallel configuration is lower than that of the antiparallel one for low temperatures, yielding a ferromagnetic interlayer coupling. For higher temperatures we indeed find a transition to an antiferromagnetic exchange coupling at $T \simeq T^*$, because the free energy of the antiparallel configuration is the lower one above this temperature.

\end{document}